\documentclass[trackchanges,twocolumn]{aastex7}

\newcommand{\Ha}{\ensuremath{\rm H\alpha}}
\newcommand{\Hbeta}{\ensuremath{\rm H\beta}}
\newcommand{\NII}{[\ion{N}{2}]}
\newcommand{\bagpipes}{\texttt{bagpipes}}
\begin{document}

\title{NGDEEP: Anchoring the $0.5 < z < 3.5$ Star-Forming Main Sequence at Low Stellar Masses with Ultradeep JWST/NIRISS WFSS}

\author[orcid=0000-0002-6913-8580, gname=Claire, sname=Bolda]{Claire Bolda}
\affiliation{Department of Physics and Astronomy, Texas A\&M University, College Station, TX, 77843-4242, USA}
\affiliation{George P. and Cynthia Woods Mitchell Institute for Fundamental Physics and Astronomy, Texas A\&M University, College Station, TX, 77843-4242, USA}
\email[show]{boldaclaire@tamu.edu}  

\author[orcid=0000-0000-0000-0002,gname=Casey, sname=Papovich]{Casey Papovich} 
\affiliation{Department of Physics and Astronomy, Texas A\&M University, College Station, TX, 77843-4242, USA}
\affiliation{George P. and Cynthia Woods Mitchell Institute for Fundamental Physics and Astronomy, Texas A\&M University, College Station, TX, 77843-4242, USA}
\email[hide]{papovich@tamu.edu}
\email{papovich@tamu.edu}

\author[orcid=0000-0001-9495-7759, gname=Lu,sname=Shen]{Lu Shen}
\affiliation{Department of Physics and Astronomy, Texas A\&M University, College Station, TX, 77843-4242, USA}
\affiliation{George P. and Cynthia Woods Mitchell Institute for Fundamental Physics and Astronomy, Texas A\&M University, College Station, TX, 77843-4242, USA}
\email{lushen@tamu.edu}

\author[orcid=0000-0001-7151-009X, gname=Cleri,sname=Nikko]{Nikko J. Cleri}
\affiliation{Department of Astronomy and Astrophysics, The Pennsylvania State University, University Park, PA 16802, USA}
\affiliation{Institute for Computational and Data Sciences, The Pennsylvania State University, University Park, PA 16802, USA}
\affiliation{Institute for Gravitation and the Cosmos, The Pennsylvania State University, University Park, PA 16802, USA}
\email{cleri@psu.edu}

\author[0000-0003-3466-035X]{{L. Y. Aaron} {Yung}}
\affiliation{Space Telescope Science Institute, 3700 San Martin Drive, Baltimore, MD 21218, USA}
\affiliation{Institute of Astronomy and Department of Physics, National Tsing Hua University, Hsinchu 30013, Taiwan}
\email{yung@phys.nthu.edu.tw}

\author[0000-0001-8489-2349]{Vicente Estrada-Carpenter}
\affiliation{School of Earth and Space Exploration, Arizona State University, Tempe, AZ 85287, USA}
\affiliation{Beus Center for Cosmic Foundations, Arizona State University, Tempe, AZ 85287, USA}
\email{vestrad9@asu.edu}

\author[0000-0001-8519-1130]{Steven L. Finkelstein}
\affiliation{Department of Astronomy, The University of Texas at Austin, Austin, TX, USA}
\affiliation{Cosmic Frontier Center, The University of Texas at Austin, Austin, TX, USA}
\email{stevenf@astro.as.utexas.edu}

\author[0000-0002-6610-2048, gname=Anton, sname=Koekemoer]{Anton M. Koekemoer}
\affiliation{Space Telescope Science Institute, 3700 San Martin Drive, 
Baltimore, MD 21218, USA}
\email{koekemoer@stsci.edu}

\author[0000-0003-3382-5941, gname=Nor, sname=Pirzkal]{Nor Pirzkal}
\affiliation{ESA/AURA, USA}
\email{pirzkal@stsci.edu}

\author[0000-0003-3903-6935]{Stephen M.~Wilkins} 
\affiliation{Astronomy Centre, University of Sussex, Falmer, Brighton BN1 9QH, UK}
\affiliation{Institute of Space Sciences and Astronomy, University of Malta, Msida MSD 2080, Malta}
\email{S.Wilkins@sussex.ac.uk}

\collaboration{all}{The NGDEEP collaboration}

\begin{abstract}
Here we present deep JWST/NIRISS wide-field slitless spectroscopy (WFSS) that covers \Ha\ and \Hbeta\ for galaxies at $0.5 < z < 3.5$ in the NGDEEP survey. The NGDEEP NIRISS observations detect emission line fluxes from galaxies in the Hubble Ultra Deep Field to $\mathrm{m_{F150W}} \lesssim$ 28 mag, enabling studies of star-formation based on \Ha\ (or \Hbeta, if necessary) in low-mass galaxies with $\mathrm{\log(M_*/M_{\odot}) \gtrsim 7}$ at $z = 0.5$ and $\mathrm{\log(M_*/M_{\odot}) \gtrsim 8}$ at $z = 3.5$. For 284 NGDEEP galaxies with $>3\sigma$ detections of \Ha\ or \Hbeta, we determine dust-corrected star-formation rates, $\mathrm{SFR_{\Ha}}$, via flux measurements of these Balmer lines and compare them to the galaxies' stellar masses to investigate the properties of the star-forming main sequence (SFMS) on short, $\sim10$~Myr timescales across three redshift bins. When we contrast the intrinsic scatter in the \Ha--based SFMS to that of a SFMS comprised of UV--based SFRs, which trace star formation within the past $\sim 100$~Myr, we find that the \Ha--based SFMS's intrinsic scatter ($\approx0.35-0.5$~dex over $0.5 < z < 3.5$) is $\approx0.2-0.4$~dex higher than the UV--based SFMS's intrinsic scatter at $z < 2.5$. We argue that brief ($\lesssim10$~Myr) bursts of star formation likely account for this disparity in SFMS scatter. Theoretical predictions from the FIRE simulations underestimate the scatter we observe in the \Ha--based SFMS at $1.5 < z < 2.5$ and $\mathrm{\log(M_*/M_{\odot}) < 9}$ by up to $\approx0.25$~dex, which suggests that low-mass galaxies have burstier star-formation histories than expected on $\lesssim10$~Myr timescales.

\end{abstract}

%% You can use the \uat command to link your UAT concepts back its source. #up to 6


\section{Introduction} 
\label{sec:intro}

%The astronomy community has dedicated a tremendous amount of effort towards gauging the star-formation histories and variability of galaxies in diverse stellar mass and redshift regimes, as models of galaxy formation and evolution attribute the cause of observed relations between the star-forming properties of galaxies and other galaxy physical properties to the baryon cycle, which converts gas that flows into a galaxy from the circumgalactic medium and beyond into stars that subsequently reintroduce gas into the circum- and inter-galactic media through stellar and supernovae winds \citep{Somerville_Dave_2015}. 

Historically, probing star-formation in low-mass galaxies ($\mathrm{\log(M_*/M_{\odot}) \lesssim 9}$) outside of the local universe ($z \gtrsim 1$) has proved unfeasible because of their perceived faintness, which diminishes at increasing $z$, thereby necessitating groundbreaking instrumental capabilities to detect them. An inevitable consequence of low-mass galaxies' elusiveness pertains to the galaxy star-forming main sequence (SFMS), a correlation that the astronomy community has uncovered between galaxies' star-formation rates (SFRs) and their stellar masses across a vast regime of stellar masses, star-formation rates, and cosmic ages \citep[e.g.,][]{Brinchmann_2004, Daddi_2007, Whitaker_2014, Shivaei_2015, Clarke_2025, Cole_2025, Merida_2026, Simmonds_2025}. Up until the comissioning of the James Webb Space Telescope (JWST), the SFMS remained entirely unexplored in the $\mathrm{\log(M_*/M_{\odot}) \lesssim 9}$ stellar mass regime at early times. 

Although anchoring the SFMS at low stellar masses is undoubtedly a formidable endeavor, even in the era of JWST, it creates invaluable new observational constraints on the morphology of the SFMS and the scatter within it, which can test predictions regarding central tenets of galaxy evolution such as formulations of feedback in galaxies \citep{Sparre_2017}. Theoretical models expect low-mass galaxies to undergo intense bursts of star formation across cosmic time due to multiple potential factors, such as: dark matter halos that are not massive enough to retain stellar feedback which could otherwise be recycled into new stars; an insufficient number of star-forming regions to maintain a constant star-formation rate; and at $z \gtrsim 1$, the outpacing of stellar feedback from supernovae by the birth of new stars \citep[e.g.,][]{Sparre_2017, Faucher_Giguere_2018, Furlanetto_2022, Hopkins_2023}. During bursts of star-formation, low-mass galaxies rise above the main sequence, and conversely, they can fall back to the main sequence or even below it in between bursts \citep{Faucher_Giguere_2018, Caplar_Tachella_2019}. As a result, models forecast that the SFMS is characterized by higher scatter at low stellar masses, especially when the SFRs involved are sensitive to recent star formation (within the past $\sim 5-10$~Myr), like SFRs estimated from \Ha\ emission \citep{Kennicutt_2012, Sparre_2017, Caplar_Tachella_2019}. By measuring the amount of scatter that exists in the \Ha\ SFMS at low stellar masses, we can gauge the intensity of feedback in low-mass galaxies in comparison to predictions, and examine how the intensity evolves as galaxies grow more massive and at increasing redshifts. Additionally, by comparing the \Ha\ SFRs of low-mass galaxies to SFRs that trace star formation on longer, $\sim$100~Myr timescales \citep{Kennicutt_2012}, such as SFRs determined observationally from the rest-UV continuum, we can quantify how often low-mass galaxies experience ``bursty'' episodes of star formation \citep{Weisz_2012,Shivaei_2015}.  

Thus far, spectroscopic observations at low redshifts ($z \lesssim 1$) have verified the ubiquity of bursty star-formation histories in galaxies with stellar masses as low as  $\mathrm{\log(M_*/M_{\odot}) \lesssim 7}$ \citep{Weisz_2012, Emami_2019}. At $z \gtrsim 1$, JWST has recently empowered photometric studies of star formation to delve to lower galaxy stellar masses than previously possible \citep[e.g.,][]{Clarke_2025, Cole_2025, Perry_2025, Simmonds_2025, Di_Cesare_2026, Merida_2026, Navarro_Carrera_2026}. Namely, \citet{Merida_2026}'s photometric analysis of the SFMS based on SED-derived, 100-Myr averaged SFRs achieved a stellar mass completeness level as low as $7.6 < \mathrm{\log(M_{*,lim}/M_{\odot}) < 8.8}$ over $1 < z < 9$ via JWST/NIRCam observations, and, in a similar fashion, \citet{Navarro_Carrera_2026}'s inquiry into variable star formation through \Ha\ and UV SFRs inferred from NIRCam photometry reached limiting stellar masses of $\mathrm{6.75 < \log(M_{*,lim}/M_{\odot}) < 7.25}$ over $2 < z < 6$. Meanwhile, spectroscopic analyses with JWST/NIRSpec of star-formation at similar redshifts have not yet managed to breach a stellar mass completeness threshold of $\mathrm{\log(M_*/M_{\odot}) \sim 8-8.5}$ \citep[e.g.,][]{Clarke_2025, Perry_2025}. Even though photometric and spectroscopic observations with JWST have already begun to provide insight into the star-formation histories of low-mass galaxies at early times, which do appear to be punctuated by bursts more frequently than higher mass galaxies as expected \citep[e.g.,][]{Clarke_2025, Perry_2025, Simmonds_2025}, deeper spectroscopic studies are still requisite to form a comprehensive view of star formation activity in galaxies with stellar masses as low as $\mathrm{\log(M_*/M_{\odot}) = 7}$ across cosmic time via direct, observationally determined SFRs. 

Here we extend investigations of the galaxy SFMS and bursty star-formation histories to $\mathrm{\log(M_*/M_{\odot}) \sim 7}$ at $z=0.5$ and $\mathrm{\log(M_*/M_{\odot}) \sim 8}$ at $z=3.5$ through JWST/NIRISS grism observations of the Hubble Ultra Deep Field (HUDF) from the Next Generation Deep Extragalactic Exploratory Public (NGDEEP) Survey \citep[PIs: S. Finkelstein, C. Papovich, and N. Pirzkal;][]{Bagley_2024}, as well as a combination of legacy optical imaging from HST/ACS and infrared imaging from JWST/NIRCam. The 1 -- 2.5~\micron\ JWST/NIRISS observations encompass Balmer emission lines such as \Ha\ and \Hbeta\ for spectroscopically detected NGDEEP galaxies at $0.5 < \mathrm{z_{grism}} < 3.5$, by means of which we can determine dust-- and metallicity--corrected \Ha\ SFRs. In addition, we estimate stellar masses and determine UV-based SFRs for the NGDEEP galaxies in this redshift range based on the previously mentioned HST and JWST photometry, which spans the rest-frame UV to near-IR portions of their spectral energy distributions. Opportunely, the redshift range we focus on in this paper, $0.5 < \mathrm{z_{grism}} < 3.5$, coincides with the era of the universe during which the amount of galaxies actively forming stars was maximal \citep{Madau_Dickinson_2014}, rendering it an ideal time-frame to study the star-forming properties of low-mass galaxies and the evolution of their star-formation activity as they grow. 

We elaborate on our $0.5 < \mathrm{z_{grism}} < 3.5$ NGDEEP sample, including its reduction, SED modeling, selection, and star-formation rate calculations, in Section \ref{section:data}. Presented in Section \ref{sec:results} are the parameters of our galaxy sample's \Ha-- and UV--based SFMSs, along with the ratios between its \Ha-- and UV--based SFRs, which we compare to previous literature measurements and theoretical predictions in Section \ref{sec:dis}. The fraction of galaxies in our sample that are actively forming stars is also considered in Section \ref{sec:dis}. 

All of the analysis presented in this paper was performed assuming the \citet{Planck_2018} cosmological model, which consists of a $\Lambda$CDM cosmology with the following parameters: $\Omega_m = 0.31$, $\Omega_{\Lambda} = 0.69$, and $H_0 = 67.7$ km~s$^{-1}$~Mpc$^{-1}$. For our galaxy metallicity estimates, we adopt the \citet{Asplund_2009} solar oxygen abundance of 12 + $\mathrm{\log(O/H)} = 8.69$ and the \citet{Steiger_Zurbuchen_2016} absolute solar metallicity of $0.0196 \pm 0.0014$. 

\section{Data and Sample}\label{section:data}

Our observations stem from extraordinarily deep, wide-field slitless spectroscopic (WFSS) and imaging surveys of the Hubble Ultra Deep Field \citep[HUDF;][]{Beckwith_2006} conducted during JWST Cycle 1, including WFSS observations from the Near-Infrared Imager and Slitless Spectrograph \citep[JWST/NIRISS,][]{Doyon_2023} as part of the  Next Generation Deep Extragalactic Exploratory Public Survey\footnote{All of the JWST observations presented in this paper are available on the Mikulski Archive for Space Telescopes (MAST) at the Space Telescope Science Institute. The NGDEEP oservations can be found on MAST at: https://doi.org/10.17909/wkee-cc37} \citep[NGDEEP; PID 2079; PIs: S. Finkelstein, C. Papovich, and N. Pirzkal;][]{Bagley_2024} and imaging from the Near-Infrared Camera (NIRCam) taken as part of the JADES\footnote{The JADES observations used in this paper are available on MAST at: https://doi.org/10.17909/8tdj-8n28} \citep{Rieke_2023, Bunker_2024, Eisenstein_2026} and JEMS\footnote{The JEMS observations involved in this paper are available on MAST at: https://doi.org/10.17909/fsc4-dt61} \citep{Williams_2023} surveys. Whereas our NIRISS observations only implement three broadband filters, F115W, F150W, and F200W, NIRCam coverage from the surveys mentioned above exists for our spectroscopically detected NIRISS sources in fourteen medium-band and and broadband filters, spanning a total observed frame wavelength range of $\sim$1 -- 5~\micron. In the rest-frame UV, we include HST/ACS imaging across four bands with a cumulative observed wavelength range of $\approx$0.4--0.85~$\mu$m from the Cosmic Assembly Near-infrared Deep Extragalactic Legacy Survey  \citep[CANDELS;][]{Grogin_2011,Koekemoer_2011}, which is another program that targeted the HUDF as well. 

We provide a brief summary of the observations and their reduction here, but we refer the reader to \citet{Bagley_2024} for the full details of the NGDEEP survey design and to \citet{Shen_2024, Pirzkal_2024, Shen_2025} for further specifics regarding the data processing. With the JWST Calibration Pipeline \citep{Bushouse_2026} and the grism redshift and line analysis software \texttt{grizli} \citep{Brammer_2023}, \citet{Shen_2025} reduced the NIRISS observations presented here into 1- and 2-D spectra. Using \texttt{grizli}, \citet{Shen_2025} fit the reduced NIRISS spectra with stellar population synthesis models from FSPS \citep{Conroy_Gunn_2009, Conroy_Gunn_2010} to ascertain their redshifts and measure their emission line fluxes. Throughout this work, we adopt these redshifts and emission line fluxes, with which we determine \Ha\ star-formation rates for the NGDEEP sample. To learn more about the \NII/\Ha\ ratios that \citet{Shen_2025} assumed in their estimation of \Ha\ fluxes for the NGDEEP sample based on the indistinguishable, \NII\ + \Ha\ profiles which NIRISS detects, please see \citet{Shen_2025}.

In an effort to gauge certain properties of the NGDEEP galaxies that NIRISS observed, such as stellar mass and SFRs averaged over 100 Myrs, \citet{Shen_2025} performed SED fitting on the JWST/NIRCam and HST/ACS photometry associated with each galaxy through the CIGALE (Code Investigating GALaxy Emission) software \citep{Boquien_2019, Yang_2020, Yang_2022}. Here we apply the resulting best-fit parameters from \citet{Shen_2025}'s SED modeling for all galaxies in our analysis, except for some galaxies with strong contamination from nearby sources in their 2D spectra that \texttt{grizli} had initially missed. For these galaxies, we masked their contamination manually; reprocessed the affected spectra with \texttt{grizli} following the same procedure as in \cite{Shen_2025}, which included re-measuring their redshifts and emission line fluxes; and repeated their SED modeling using the same star-formation history model and parameter constraints as \citet{Shen_2025}. Our study achieves one of NGDEEP's primary aims: to constrain the scatter in the \Ha\ SFMS at stellar masses of $\mathrm{\log(M_*/M_{\odot}) \gtrsim 7}$, which in turn helps gauge the ``burstiness'', or variability, of star-formation at these low stellar masses \citep{Bagley_2024}.

\subsection{Sample Selection} \label{subsec:sample}

Our parent sample consists of all spectroscopically detected sources in the NGDEEP JWST/NIRISS galaxy catalog. We construct our initial sample by drawing sources from the parent sample with coverage of \Ha\ in the NIRISS WFSS at redshifts $0.5 < \mathrm{z_{grism}} < 2.4$ and \Hbeta\ from $1.1 < \mathrm{z_{grism}} < 3.5$ in our F115W, F150W, and F200W filters. For each source in our initial sample, we visually inspect its 1D and 2D spectrum, and remove objects characterized by any of the following traits: the lack of a continuum or the presence of a low signal-to-noise continuum; a lack of emission or absorption features or spectral breaks; and contamination of the spectrum by a nearby, bright source for which we cannot correct. Additionally, we remove 20 X--ray--detected AGN from our sample that \citet{Shen_2025} previously identified by comparing the sky coordinates of the full NGDEEP sample to those of AGN in \citet{Lyu_2022}'s observations from the Chandra X-ray Observatory. We then create a final sample limited to those galaxies with stellar masses above our 80\% empirical completeness threshold, which is described below in Section \ref{subsubsec:mass_completeness}. Our final sample, then, includes 414 galaxies that satisfy the above conditions; the stellar masses and redshifts of these galaxies are displayed in Figure \ref{fig:mstar_zgrism} below.

\begin{figure*}[h!]
\centering
\includegraphics[scale=0.5]{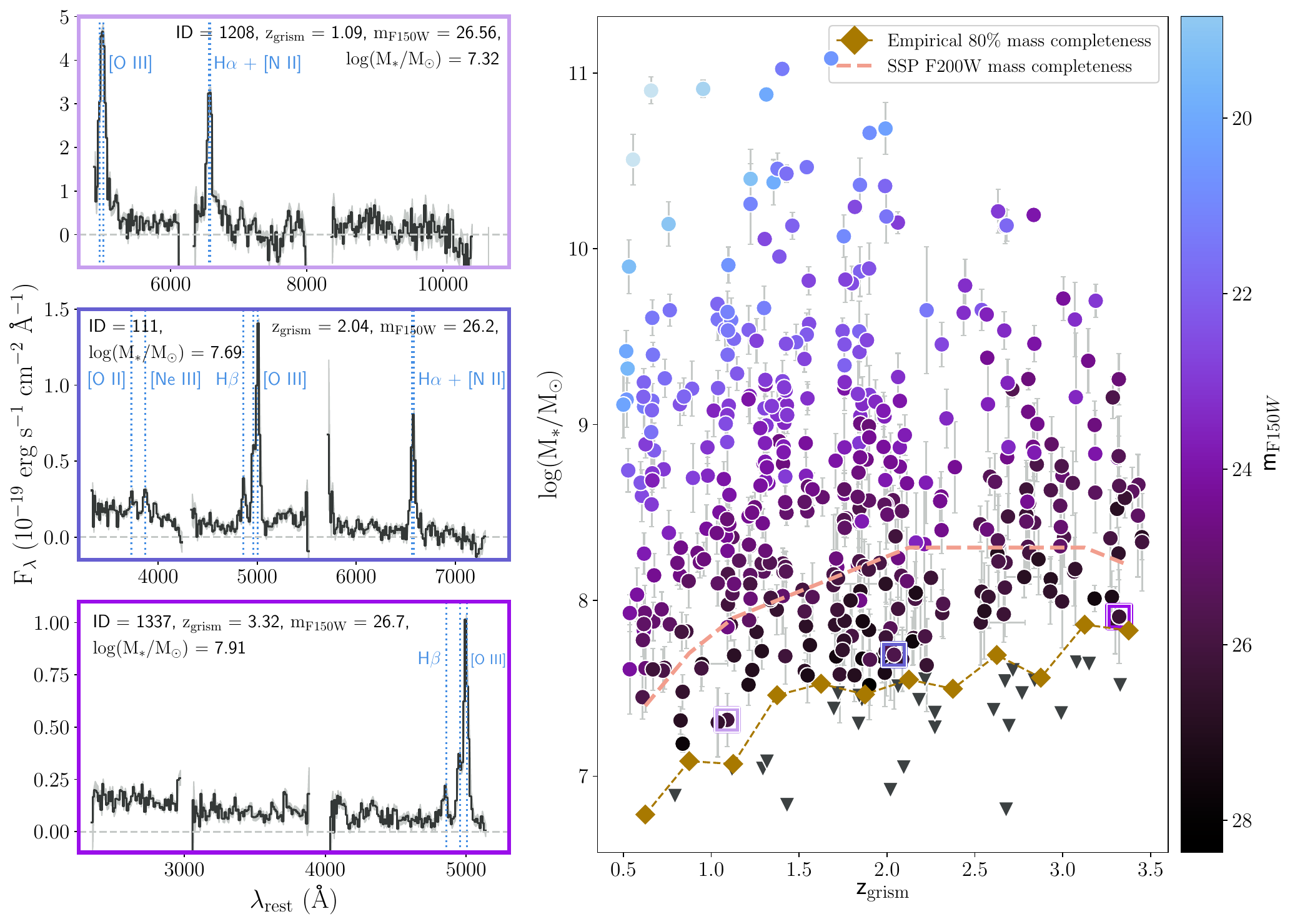}
\caption{Here, the right panel illustrates the stellar masses of our final sample plotted as a function of redshift, with brighter galaxy magnitudes represented as lighter colors and fainter magnitudes by darker colors. Gold diamonds delineate the 80\% mass completeness level for our sample, which was estimated using the continuum flux signal-to-noise ratios of our galaxy sample. Likewise, the pink dashed line signifies the mass completeness level for a simple stellar population formed at $z=4$ in the F200W band. The determination of both mass-completeness levels are discussed further in Section \ref{subsubsec:mass_completeness}. Black triangles represent spectroscopically confirmed galaxies of good spectral quality that lie beneath our redshift-dependent mass completeness level. The lavender, periwinkle, and violet squares highlight three galaxies at different redshifts near the stellar-mass limit.  The 1D NIRISS spectrum of each of these galaxies is shown in the left-hand panels with the same color code. \label{fig:mstar_zgrism}}
\end{figure*}

\subsubsection{Mass Completeness} \label{subsubsec:mass_completeness}

To account for potential non-detections of faint, low-mass galaxies, we determine a redshift-dependent, empirical stellar--mass limit for our sample following a similar approach to \citet{Tomczak_2014}, which involves two different methods. First, we ascertain an 80\% mass completeness level by simultaneously decreasing the signal-to-noise ratio of every spectroscopically confirmed galaxy's\footnote{including galaxies that we ultimately exclude from the final sample due to their poor spectral quality} continuum flux in each band, which is measured by \texttt{grizli} for every galaxy's spectrum, and stellar mass by $0.25 \ \mathrm{dex}$ in redshift bins of $\Delta z = 0.25$ and stellar mass bins of $\mathrm{log(M_*/M_{\odot})} = 0.25$ until 20\% of the galaxies' continuum signal-to-noise ratios drop below 5 in any band and each redshift bin. We then record the 80\% stellar-mass completeness limit as the median stellar mass of the galaxies in this redshift bin, and we exclude both star-forming and quiescent galaxies with stellar masses below this limit from our final sample. At redshifts of $\mathrm{z_{grism} = [0.5, 2, 3.5]}$, we find mass completeness limits of $\mathrm{log(M_*/M_{\odot})} = [6.8, 7.5, 7.8]$ for our sample. 

Secondly, as in \citet{Tomczak_2014}, we compare our sample to a simple stellar population (SSP) formed in a burst at $z = 4$ with a fixed metallicity of $Z_* = 0.01$ (or $Z_* = 0.5Z_{\odot}$), which we simulated using \bagpipes\ assuming the \citet{Bruzual_2003} models. Each model is evolved forward in time (e.g., decreasing redshift) from $z=4$, such that the SSP galaxies have observed redshifts ranging from $0.5 < \mathrm{z_{obs}} < 3.5$ in increments of $\mathrm{\Delta z_{obs} = 0.25}$. For each model galaxy in the SSP, we simultaneously lower its stellar mass and flux in a given band by $0.1~$dex until its flux drops below 80\% of the maximum continuum flux in that band for NGDEEP's spectroscopically detected galaxies at $0.5 < \mathrm{z_{obs}} < 3.5$. We then record the decreased stellar mass of each model galaxy as the SSP mass completeness level at the galaxy's redshift for a given band, producing a redshift-dependent mass-completeness curve over $0.5 < \mathrm{z_{obs}} < 3.5$ like the one shown in red in Figure \ref{fig:mstar_zgrism} for the F200W band. Our ability to detect star-forming and quiescent galaxies hinges not only upon this mass-completeness limit, however, but also on our redshift-dependent SFR limits, which we discuss further in Section \ref{ssec:sfms}.

\subsection{SFR Measurements}\label{subsec:sfr}

In this work, we compare the SFRs and SFMS from the \Ha\ and \Hbeta\ lines to the 100-Myr averaged SFRs derived from SED fitting with \texttt{cigale}. Through this comparison, we can trace star-formation on different timescales because the \Ha\ and \Hbeta\ emission lines probe the birthrates of the most massive stars (with lifetimes of $\sim5-10$~Myr), and conversely, the SED-derived SFRs are based primarily on the rest-UV continuum, which is sensitive to the formation of stars within the previous $\sim 100$~Myr \citep{Kennicutt_2012}. However, within our $0.5 < z_{\mathrm{grism}} < 3.5$ sample, \Ha\ and \Hbeta\ are only simultaneously covered by NGDEEP's JWST/NIRISS grism observations from $z_{\mathrm{grism}} \approx 1.1$ in the F115W filter to $z_{\mathrm{grism}} \approx 2.4$ in the F200W filter. For this reason, and because \Ha\ and/or \Hbeta\ unfortunately sometimes fall into the detector gaps for certain galaxies in our sample, we cannot determine dust-corrected \Ha\ star-formation rates for all sample galaxies in a uniform manner. We discuss the two different dust corrections we apply to our sample galaxies, depending on which Balmer features are available to us, below.

For 196/414 (or $\approx47$\%) of the galaxies in our sample,  \Ha\ and \Hbeta\ are both detected in emission at $1.1 < z_{\mathrm{grism}} < 2.4$. We correct these galaxies' observed \Ha\ fluxes for dust following the \citet{Calzetti_2000} attenuation law. Whenever possible, we implement the reddening of the nebular lines determined observationally from the Balmer decrement, $\mathrm{E(B-V)_{BD}}$, in these corrections. In some cases, though, we find that the signal-to-noise ratio of the nebular reddening is low ($\lessapprox 3$) due to marginal Balmer-line detections. Because of this, we convert the reddening of the stellar continuum estimated from SED fitting with \texttt{cigale}, $\mathrm{E_s(B-V)_{SED}}$, into a nebular reddening estimate ($\mathrm{E(B-V)_{SED,neb}}$) by dividing it by 0.44, following Equation 9 in \citet{Calzetti_2000}, and compare it to $\mathrm{E(B-V)_{BD}}$ for all 196 galaxies. If $\mathrm{E(B-V)_{SED,neb}}$ has a higher signal-to-noise ratio than the observed $\mathrm{E(B-V)_{BD}}$, we substitute it for the nebular reddening estimate when applying dust corrections.

For the 102/414 (or $\approx25$\% of) galaxies in our sample at $z_{\mathrm{grism}} < 2.4$ that have detected \Ha\ emission but lack measurements of \Hbeta, either because \Hbeta\ is not covered (at $z_{\mathrm{grism}} < 1.1$) or falls into a detector gap, we likewise calculate $\mathrm{E(B-V)_{SED,neb}}$ from \texttt{cigale}'s estimate of the stellar reddening as described above and use it to correct the \Ha\ fluxes for dust. At $z_{\mathrm{grism}} > 2.4$, however, \Hbeta\ is the only Balmer line available for characterizing star-formation. For the 76 galaxies in our sample at $z_{\mathrm{grism}} > 2.4$ with detected \Hbeta\ emission and the 7 galaxies at $1.1 < z_{\mathrm{grism}} > 2.4$ that have detected \Hbeta\ emission but are missing \Ha\ emission due to detector gaps (in total, $\approx20$\% of our sample), we apply a correction of 2.87 to convert their observed \Hbeta\ fluxes into \Ha\ line flux estimates assuming Case-B recombination. We then follow the same steps outlined above, using $\mathrm{E(B-V)_{SED,neb}}$, to estimate dust-corrected \Ha\ fluxes for these galaxies. The remaining 33 galaxies in (or $\approx8$\% of) our sample exhibit no Balmer--line--emission whatsoever, either because they are quiescent or because we only detect their continuum; regardless, we cannot derive a star-formation rate for these galaxies based on our spectroscopic observations.

Because 283/414 (or $\approx 68$\% of) galaxies in our sample possess stellar masses of $\mathrm{\log(M_*/M_{\odot}) < 9}$, and because low-mass galaxies tend to have subsolar metallicities \citep[e.g.,][]{Tremonti_2004, Henry_2013, Guo_2016}, especially at $z \gtrsim 1$ \citep[e.g.,][]{Erb_2006, Sanders_2021}, we must gauge the metallicities for our sample and then assess whether the \Ha\ SFR calibration from \cite{Kennicutt_2012}, which applies to galaxies with solar metallicities, is still appropriate for our sample. Although we do not have consistent coverage of strong lines with which we can directly estimate metallicities for our entire sample, we can obtain reasonable metallicity estimates for our sample galaxies by extrapolating from the \citet{Sanders_2021} mass-metallicity, power-law relationships at $z = 2.3$ and $z=3.3$. Our sample's resulting median metallicity is $Z_* = 0.25Z_{\odot} \approx 0.005$. On account of multiple studies which show that the \Ha\ SFR calibration depends on metallicity \citep{Wilkins_2019, Korhonen_Cuestas_2025}, the canonical \cite{Kennicutt_2012} SFR calibration based on the assumption of Solar metallicity would substantially overestimate the \Ha\ SFRs for many galaxies in our sample. Therefore, we use the \citet{Korhonen_Cuestas_2025} SFR calibration, which modifies the constant conversion factor from the \citet{Kennicutt_2012} relation to a variable conversion factor that increases rapidly with decreasing metallicity over the metallicity range $0.001 < Z_* < 0.02$, to convert our sample galaxies' dust-corrected \Ha\ fluxes into dust-- and metallicity--corrected \Ha\ SFRs. Given our sample's range of metallicities, $0.001 < Z_* < 0.03$, \Ha\ SFRs calculated assuming the conversion factor from \citet{Kennicutt_2012} are $0.05-0.4~$dex greater than the \Ha\ SFRs calculated using the metallicity-dependent conversion factor from \citet{Korhonen_Cuestas_2025}, with the largest difference occurring at the lowest metallicities. 

Throughout this work, we define $\mathrm{SFR_{\Ha}}$ to be the SFR measured from either the dust- and metallicity-corrected \Ha\ or \Hbeta\ emission line, as both probe star-formation on short-duration, 5--10~Myr timescales. Similarly, we define $\mathrm{SFR_{UV}}$ to be the dust-corrected SED-derived SFRs averaged over 100 Myrs from \texttt{cigale}.

\section{Results}\label{sec:results}

\subsection{The Star-Forming Main Sequence and Its Evolution Across $\mathit{0.5 < z_{grism} < 3.5}$}\label{ssec:sfms}

\begin{figure*}[h!]
\includegraphics[width=\linewidth]{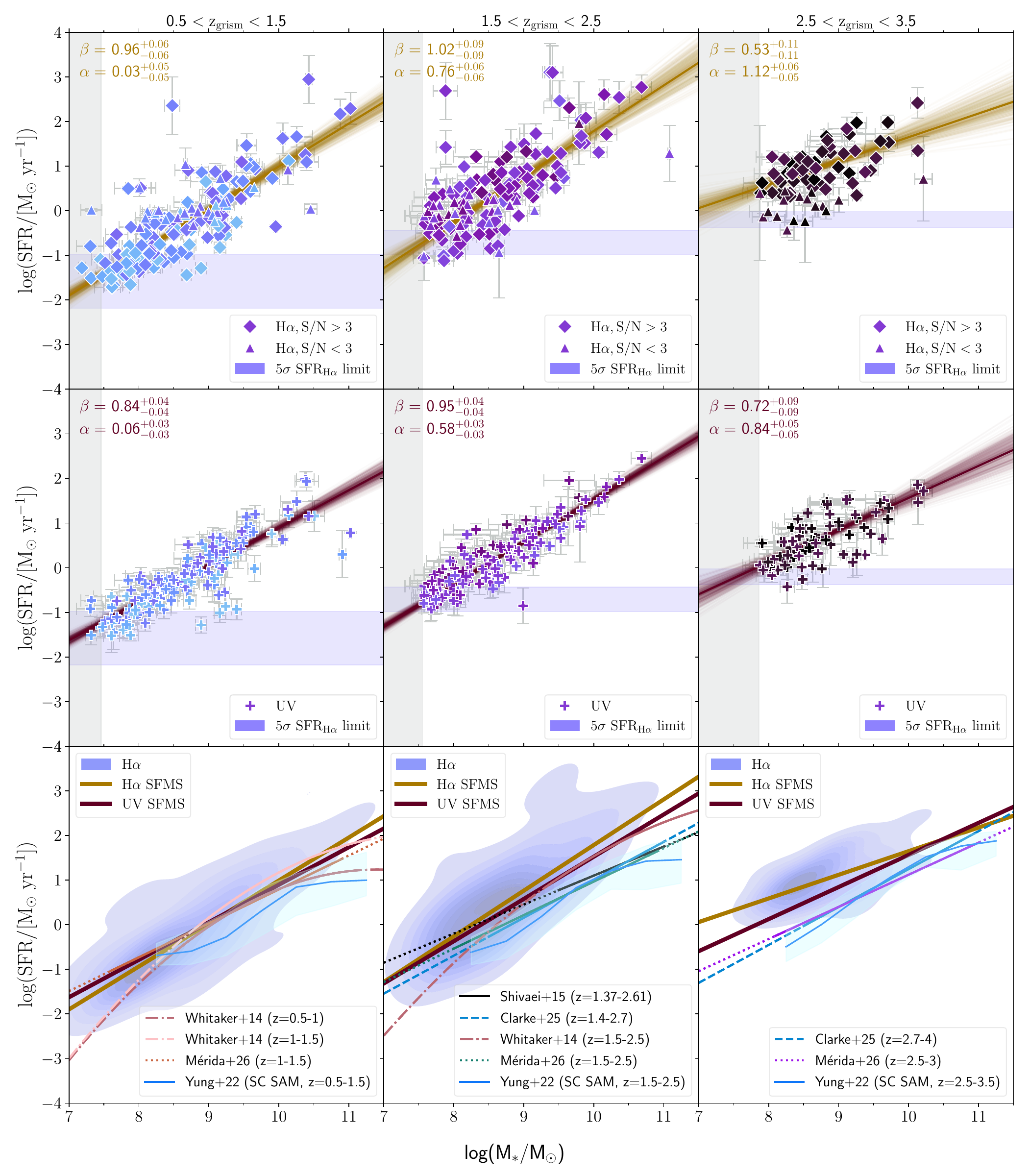}
\caption{NGDEEP SFMS relations derived from the 
%
% Balmer-line <-- CJP commented this to keep Lu's comment, which is a good one I think. 
%
\Ha\ SFRs (top row) and UV SFRs (middle row) for our final $0.5 < \mathrm{z_{grism}} < 3.5$ NGDEEP sample.  In each row, the panels show the best-fit \Ha-- and UV--based SFMS relations from \texttt{linmix} for three redshift bins as solid bronze and burgundy lines, respectively. The diamonds and crosses represent \Ha\ SFRs with a significance of $3\sigma$ or greater and UV SFRs, respectively, and are color coded by redshift such that galaxies with lower redshifts have lighter colors (and vice versa).
Conversely, upward-pointing triangles represent \Ha\ SFRs that we detect at a significance lower than $3\sigma$. Our 5$\sigma$ $\mathrm{\log(SFR)}$ detection limit is delineated by rectangular, shaded periwinkle regions; further discussion about this detection threshold can be found in Section \ref{sec:results}.  Likewise, our redshift-dependent mass-completeness limits are shown as gray shaded contours in the top two rows. The bottom row compares our sample's $\mathrm{SFR_{\Ha}}$--$\mathrm{M_*}$ distribution (shaded periwinkle contours) and best-fit \texttt{linmix} SFMSs to previous SFMS relationships observed in the literature and predicted by the Santa Cruz semi-analytical model \citep{Yung_2022}. We discuss these comparisons further in Section \ref{sec:dis}.
\label{fig:sfms}}
\end{figure*}

To ascertain the morphology of the SFMS and the scatter within it on both $\mathrm{\sim 5-10 \ Myr}$ and $\mathrm{\sim 100 \ Myr}$ timescales for our final $0.5 < z_{\mathrm{grism}} < 3.5$ sample, we subdivide our sample into two overlapping groups of galaxies with \Ha\ SFRs and UV SFRs and then perform Bayesian linear regression on both groups' SFRs separately in redshift bins of $\Delta z = 1$ using \texttt{linmix} \citep{Kelly_2007}, which measures the slope, intercept, and intrinsic scatter of the \Ha\ and UV SFMSs.

In our analysis of the \Ha\ SFMS, we only include galaxies with $\mathrm{\log(SFR_{\Ha})}$ above a S/N threshold of 3 and above a $\mathrm{\log(SFR)}$ threshold inferred from our band-dependent 5$\sigma$ emission line detection limits of $1.2 \times 10^{-18} \ \mathrm{ergs \ s^{-1} \ cm^{-1}}$ in the F115W band, $1.3 \times 10^{-18} \ \mathrm{ergs \ s^{-1} \ cm^{-1}}$ in the F150W band, and $1.5 \times 10^{-18} \ \mathrm{ergs \ s^{-1} \ cm^{-1}}$ in the F200W band, which were previously determined by \citet{Bagley_2024}. The determination of the individual $\mathrm{\log(SFR)}$ threshold for each galaxy depends both on its redshift and the observed wavelength of the Balmer emission feature (either \Ha\ or \Hbeta) from which its \Ha\ SFR originates. In addition, for both the \Ha\ and UV SFMSs, we restrict our analysis to star-forming galaxies (i.e., we remove quiescent galaxies) by implementing a redshift-dependent $\mathrm{\log(sSFR)}$ cut of $\mathrm{\log(sSFR) > 0.2 / t_H }$, where $t_H$ represents the Hubble time, as encouraged by  \citet{Whitaker_2026} for galaxies at $z \gtrsim 1$. 

As for the the analysis of the UV-based SFMS, we do not apply any S/N cuts, but we do exclude galaxies that lie below the aforementioned sSFR cut and band-dependent SFR threshold from our analysis of the UV--based SFMS. If a galaxy is included in both SFMSs (i.e., has one or more detected Balmer emission feature), we simply use the exact same SFR threshold value to accomplish this; however, if a galaxy only has a UV--based SFR, we calculate its unique SFR threshold based on its redshift and the band in which the observed wavelength of \Ha\ lies for this galaxy up until $\mathrm{z_{grism} = 2.4}$, where the only type of Balmer emission we could possibly observe in this galaxy's spectrum is \Hbeta\ in the F200W filter. Out of the 414 galaxies in our final $0.5 < \mathrm{z_{grism}} < 3.5$ NGDEEP sample, 284 (396) galaxies are included in our analysis of the \Ha\ (UV) SFMS.

In Figure \ref{fig:sfms} we display the fitted SFMS relationships from \texttt{linmix} for the $\mathrm{SFR_{\Ha}}$ (bronze) and $\mathrm{SFR_{UV}}$ (burgundy) samples in each of the three redshift bins. Additionally, we provide the 50th percentile values for the slope $\beta$ and the intercept $\alpha$ of the linear relationships in the top left of the first two rows of Figure \ref{fig:sfms}, and likewise, we present the 50th percentile value for the intrinsic scatter $\mathrm{\sigma_{int}}$ in Table \ref{tab:sfms_scatter} below. For all three quantities, we report lower and upper uncertainties that stem from the 16th and 84th percentiles of their posterior distributions, respectively. 

\begin{deluxetable*}{c c c c c c c c c}[ht!]
\tablewidth{0pt}
\tablecaption{Scatter within the SFMS \label{tab:sfms_scatter}}
\tablehead{
\colhead{$\mathrm{\log(M_*/M_{\odot})}$} & \colhead{$\mathrm{N_{\Ha}}$} & \colhead{$\mathrm{N_{UV}}$} & \colhead{$\mathrm{\sigma_{int, \Ha}}$} & \colhead{$\mathrm{\sigma_{int, UV}}$} & \colhead{$\mathrm{\sigma_{NMAD}}$} & \colhead{$\mathrm{\sigma_{NMAD}}$} & \colhead{$\mathrm{\sigma_{NMAD}}$} & \colhead{$\mathrm{\sigma_{NMAD}}$} \\
\colhead{range} & \colhead{} & \colhead{} & \colhead{} & \colhead{} & \colhead{$\mathrm{\log(SFR_{\Ha,obs})}$} & \colhead{$\mathrm{\log(SFR_{\Ha,corr})}$} &  \colhead{$\mathrm{\log(SFR_{UV,obs})}$} & \colhead{$\mathrm{\log(SFR_{UV,corr})}$} \\
\colhead{(1)} & \colhead{(2)} & \colhead{(3)} & \colhead{(4)} & \colhead{(5)} & \colhead{(6)} & \colhead{(7)} & \colhead{(8)} & \colhead{(9)} 
}
\startdata
  \multicolumn{9}{c}{$0.5 < \mathrm{z_{grism}} < 1.5$} \\
\tablebreak
\hline
full & 124 & 169 & $0.48^{+0.04}_{-0.03}$ & $0.30^{+0.03}_{-0.02}$ & $0.47^{+0.06}_{-0.06}$ & $0.47^{+0.05}_{-0.06}$ & $0.29^{+0.03}_{-0.03}$ & $0.20^{+0.05}_{-0.04}$ \\
$<8^{*}$ & 30 & 33 & -- & -- & $0.39^{+0.10}_{-0.07}$ & $0.38^{+0.10}_{-0.07}$ & $0.25^{+0.13}_{-0.07}$ & $0.13^{+0.23}_{-0.13}$ \\
$8 - 9$ & 53 & 76 & -- & -- & $0.49^{+0.09}_{-0.09}$ & $0.49^{+0.09}_{-0.09}$ & $0.23^{+0.03}_{-0.03}$ & $0.10^{+0.08}_{-0.06}$ \\
$9 - 10$ & 33 & 50 & -- & -- & $0.45^{+0.16}_{-0.09}$ & $0.44^{+0.16}_{-0.09}$ & $0.30^{+0.08}_{-0.06}$ & $0.19^{+0.13}_{-0.09}$ \\
\hline
  \multicolumn{9}{c}{$1.5 < \mathrm{z_{grism}} < 2.5$} \\
\tablebreak
\hline
full & 103 & 142 & $0.53^{+0.05}_{-0.05}$ & $0.15^{+0.03}_{-0.03}$ & $0.65^{+0.08}_{-0.06}$ & $0.64^{+0.08}_{-0.06}$ & $0.30^{+0.03}_{-0.03}$ & $0.22^{+0.04}_{-0.03}$ \\
$<8^{*}$ & 23 & 36 & -- & -- & $0.52^{+0.19}_{-0.10}$ & $0.51^{+0.19}_{-0.10}$ & $0.29^{+0.04}_{-0.06}$ & $0.21^{+0.06}_{-0.08}$ \\
$8 - 9$ & 46 & 68 & -- & -- & $0.69^{+0.10}_{-0.12}$ & $0.69^{+0.11}_{-0.12}$ & $0.32^{+0.05}_{-0.04}$ & $0.25^{+0.06}_{-0.05}$ \\
$9 - 10$ & 29 & 33 & -- & -- & $0.58^{+0.16}_{-0.17}$ & $0.56^{+0.17}_{-0.17}$ & $0.28^{+0.05}_{-0.06}$ & $0.21^{+0.07}_{-0.08}$ \\
\hline
  \multicolumn{9}{c}{$2.5 < \mathrm{z_{grism}} < 3.5$} \\
\tablebreak
\hline
full & 57 & 85 & $0.35^{+0.04}_{-0.04}$ & $0.34^{+0.04}_{-0.03}$ & $0.40^{+0.07}_{-0.06}$ & $0.37^{+0.08}_{-0.06}$ & $0.30^{+0.06}_{-0.05}$ & $0.26^{+0.07}_{-0.05}$ \\
$8 - 9$ & 39 & 58 & -- & -- & $0.35^{+0.07}_{-0.08}$ & $0.32^{+0.08}_{-0.09}$ & $0.31^{+0.07}_{-0.07}$ & $0.28^{+0.08}_{-0.08}$ \\
$9 - 10$ & $<15^{\dagger}$ & 18 & -- & -- & -- & -- & $0.45^{+0.15}_{-0.16}$ & $0.41^{+0.17}_{-0.17}$ \\
\tableline
\enddata
\tablecomments{(1) Stellar mass range, in dex. Here, $\mathrm{M_{*,lim}(z)}$ represents the redshift-dependent $80\%$ spectroscopic mass-completeness level for our sample. \\
(2) Number of galaxies above the star-formation threshold with $\mathrm{S/N > 3}$ for $\mathrm{\log(SFR_{\Ha})}$ and $\mathrm{log(sSFR_{\Ha}) > 0.2 / t_H}$ in a given bin of stellar mass and redshift. \\ 
(3) Number of galaxies above the star-formation threshold with $\mathrm{log(sSFR_{UV}) > 0.2 / t_H}$ in a given bin of stellar mass and redshift. \\
(4) Intrinsic scatter in the \Ha\ SFMS from \texttt{linmix}. \\
(5) Intrinsic scatter in the UV--based SFMS from \texttt{linmix}. \\
(6) Observational $\mathrm{\sigma_{NMAD} \ \log(SFR_{\Ha})}$ values with uncertainties from Monte Carlo simulations. \\
(7) Observational $\mathrm{\sigma_{NMAD} \ \log(SFR_{\Ha})}$ values with uncertainties from Monte Carlo simulations that have been corrected for uncertainties on $\mathrm{log(SFR_{\Ha})}$. \\
(8) Observational $\mathrm{\sigma_{NMAD} \ \log(SFR_{UV})}$ values with uncertainties from Monte Carlo simulations. \\
(9) Observational $\mathrm{\sigma_{NMAD} \ \log(SFR_{\Ha})}$ values with uncertainties from Monte Carlo simulations that have been corrected for uncertainties on $\mathrm{log(SFR_{UV})}$. \\
$\dagger$ An insufficient number of galaxies is available to perform a Monte Carlo simulation in this bin of stellar mass and redshift. \\
$*$ Includes galaxies with stellar masses between the redshift-dependent, 80\% mass-completeness level and $\mathrm{\log(M_*/M_{\odot}) = 8}$}
\end{deluxetable*}  
    
\medskip 

\begin{figure*}[ht!]
\centering
\includegraphics[scale=0.375]{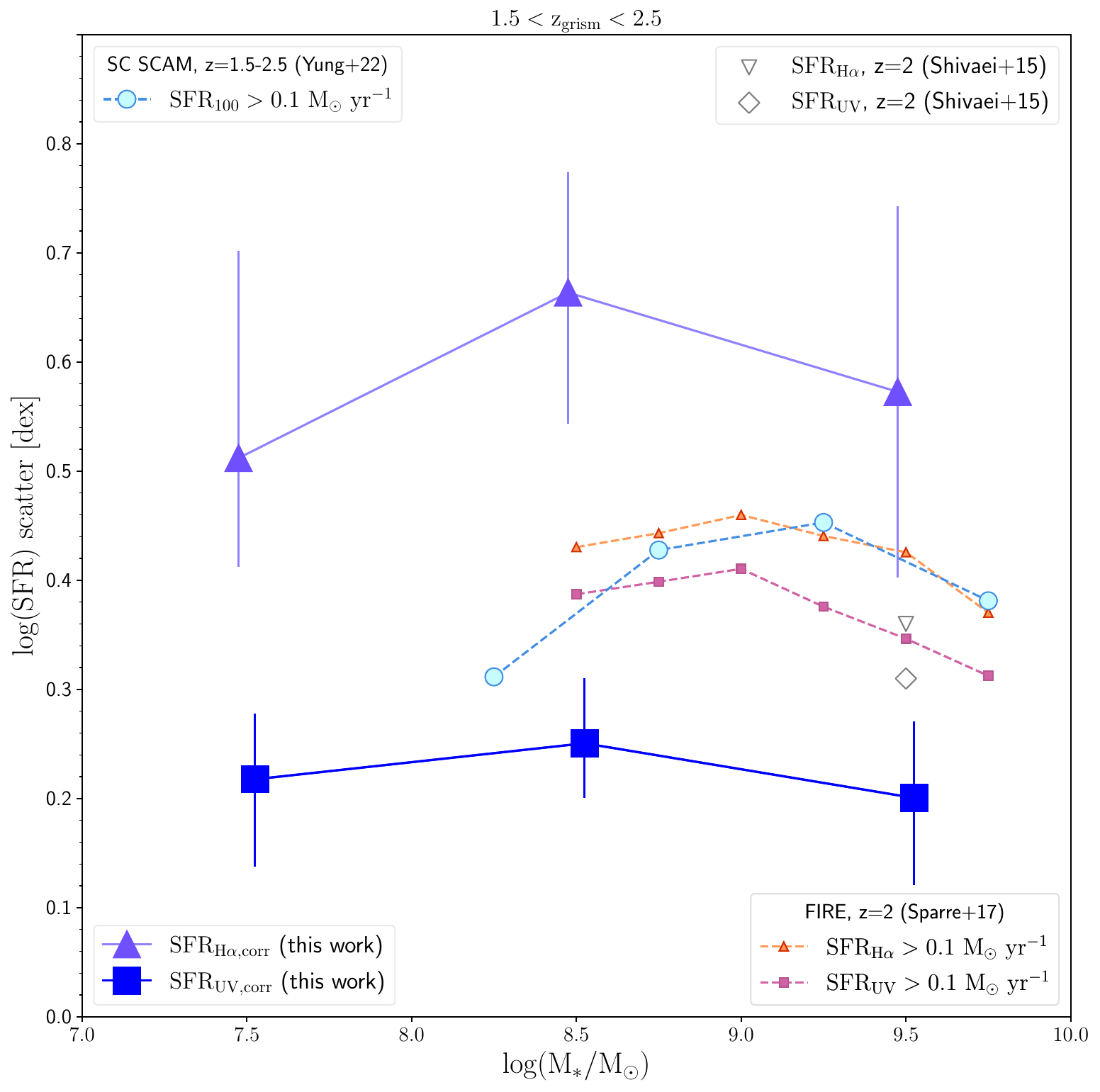}
\includegraphics[scale=0.375]{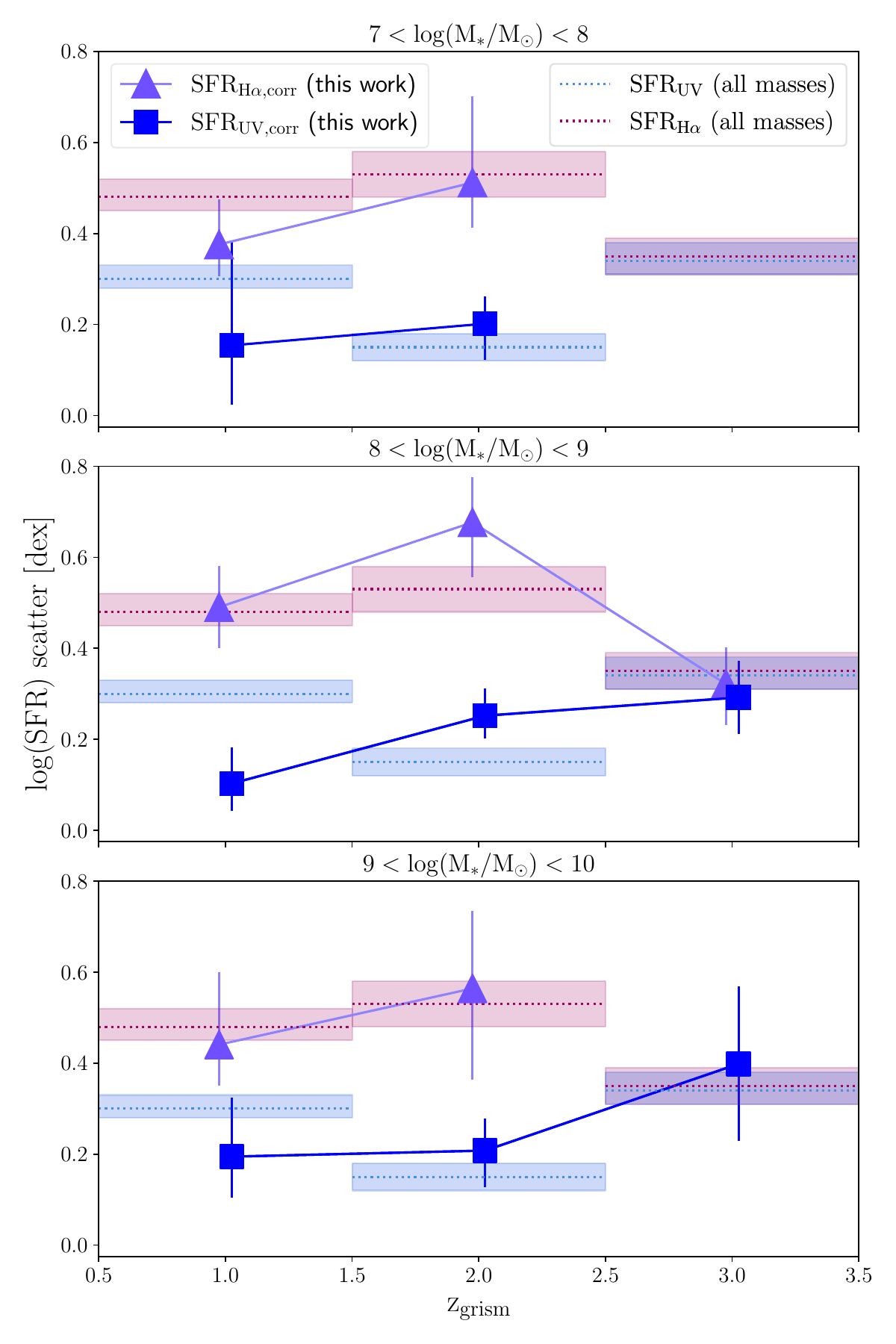}
\caption{\textit{Left}: The scatter in the \Ha-- and UV--based SFMSs at $1.5 < z < 2.5$ as a function of stellar mass. Our NGDEEP scatter measurements that are corrected for uncertainties, $\mathrm{\sigma_{NMAD} \ \log(SFR_{corr})}$, are represented here by periwinkle triangles for the \Ha--based SFMS and by blue squares for the UV--based SFMS.  The singular open triangle and diamond symbolize previous SFMS scatter measurements for \Ha\ and and UV, respectively, from MOSDEF \citep{Shivaei_2015} at similar redshifts.  Predictions of the scatter in the SFMS from FIRE at z=2 \citep{Sparre_2017} are depicted by orange triangles for \Ha\ and pink squares for UV. Likewise, light blue circles show SFMS scatter predictions involving SFRs averaged over 100-Myr timescales from the SC SAM \citep{Yung_2022}. For both simulations, the SFRs are restricted to SFR $>$ 0.1 $M_\odot$ yr$^{-1}$), which roughly corresponds to the minimum SFR we can detect in our sample at $\mathrm{1.5 < z_{grism} < 2.5}$ due to our $5\sigma$ emission line detection limit. \textit{Right:} The scatter in the SFMS as a function of redshift within three bins of stellar mass. The periwinkle triangles and blue squares are the same here as in the left panel. Shaded blue and pink regions represent the intrinsic scatter of the SFMS calculated by \texttt{linmix} across all stellar masses in three bins of redshift. In both the left and right panels of this figure, the uncertainties on the $\mathrm{\sigma_{NMAD} \ \log(SFR_{corr})}$ measurements originate from a Monte Carlo simulation with 400 iterations, which is discussed further in Section \ref{sec:results}.
\label{fig:scatter}}
\end{figure*}

\subsection{Quantifying the Dependence of the SFMS Scatter on Stellar Mass and Redshift}\label{subsec:results_bins}

To estimate the scatter along the star-forming main sequence as a function of both redshift and stellar mass, we first subdivide the \Ha\ and UV-based SFMSs for our sample into three redshift bins and three stellar mass bins over the stellar-mass range $7 < \mathrm{\log(M_*/M_{\odot}) < 10}$, such that each SFMS contains 9 two-dimensional bins with a length of $\Delta \mathrm{\log(M_*/M_{\odot}}) = \mathrm{1 \ dex}$ and a height of $\Delta z = 1$. In each of the nine resulting bins that contains at least 15 galaxies, we compare the galaxies' $\mathrm{\log(SFRs)}$ to the appropriate best-fit SFMS determined by \texttt{linmix} for the same redshift range. We calculate the difference between the galaxies' $\mathrm{\log(SFRs)}$ and the $\mathrm{\log(SFR)}$ values of the best-fit SFMS evaluated at the stellar masses of the galaxies, or $\mathrm{\Delta (\log(SFR) - SFMS)}$, following \citep{Shivaei_2015}, and take the mean absolute deviation $\mathrm{\sigma_{NMAD}}$ of those differences to be the observed scatter in the SFMS within a given bin of stellar mass and redshift, which we refer to hereafter as $\mathrm{\sigma_{NMAD} \log(SFR_{obs})}$. Altogether, we find that 7 bins in the \Ha\ SFMS and 8 bins in the UV-based SFMS have a sufficient number of galaxies for us to measure $\mathrm{\sigma_{NMAD} \ log(SFR_{obs})}$ robustly. Across our entire redshift range, there are less than 20 galaxies in our \Ha-- and UV--based  SFMSs with stellar masses above $\mathrm{\log(M_*/M_{\odot}}) \sim 10$ due to the small FOV of our NIRISS observations, which renders our sample less complete for galaxies at higher stellar masses than some studies of the SFMS at similar redshifts, such as \citet{Whitaker_2014}, \citet{Shivaei_2015}, and \cite{Merida_2026}. Therefore, we are not able to constrain the scatter in the SFMS at comparably high stellar masses as the literature. Instead, we focus on the evolution at the low-mass end of the SFMS.  

We estimate the uncertainties on the scatter for each $\Delta z = 1$, $\Delta \mathrm{\log(M_*/M_{\odot}}) = \mathrm{1 \ dex}$ bin through a Monte Carlo simulation with 400 iterations in which we select random galaxies in a given bin with replacement and evaluate the mean absolute deviation of their $\mathrm{\Delta (\log(SFR) - SFMS)}$ values, or $\mathrm{\sigma_{NMAD} \ log(SFR_{obs})}$, as explained above. In Table \ref{tab:sfms_scatter}, we report the median value of the resulting $\mathrm{\sigma_{NMAD} \ log(SFR_{obs})}$ distribution in columns (6) and (8), with lower and upper uncertainties originating from the 16th and 84th percentiles of the same distribution. However, these mean absolute deviations do not take into account the contribution of SFR uncertainties to the scatter that we measure directly from the SFMS. For this reason, we propagate the uncertainties on the random galaxies' $\mathrm{\log(SFRs)}$ through our calculation of $\mathrm{\sigma_{NMAD} \log(SFR_{obs})}$, and following the procedure outlined in \citet{Shivaei_2015}, we correct our observed SFMS scatter measurements by subtracting the median $\mathrm{\Delta (\log(SFR) - SFMS)}$ uncertainty associated with the randomly selected galaxies in the Monte Carlo simulation from the observed $\mathrm{\sigma_{NMAD} \ log(SFR_{obs})}$ measurement for the selected galaxies in quadrature. The corrected $\mathrm{\sigma_{NMAD} \ log(SFR)}$ measurements, or $\mathrm{\sigma_{NMAD} \ log(SFR_{corr})}$, for the \Ha-- and UV--based  SFMSs are available in columns (7) and (9) of Table~\ref{tab:sfms_scatter} and are exhibited in both panels of Figure \ref{fig:scatter} at various redshifts and stellar masses as well. Trends that we discern between the corrected, \Ha-- and UV--based SFMS scatter and theoretical outlooks for the scatter in the SFMS are considered in Section \ref{subsec:discussion_bins}. Additionally, we repeat the Monte Carlo simulations across the entire (or ``full") stellar-mass range of the sample above the 80\% mass completeness level in each redshift bin; the resulting observed and corrected SFMS scatter measurements are available in Table \ref{tab:sfms_scatter}. 

\subsection{The Influence of z and $\mathit{M_*}$ on the $\mathit{SFR_{\Ha}/SFR_{UV}}$ Relation}

\begin{figure*}[ht!]
\includegraphics[width=\linewidth]{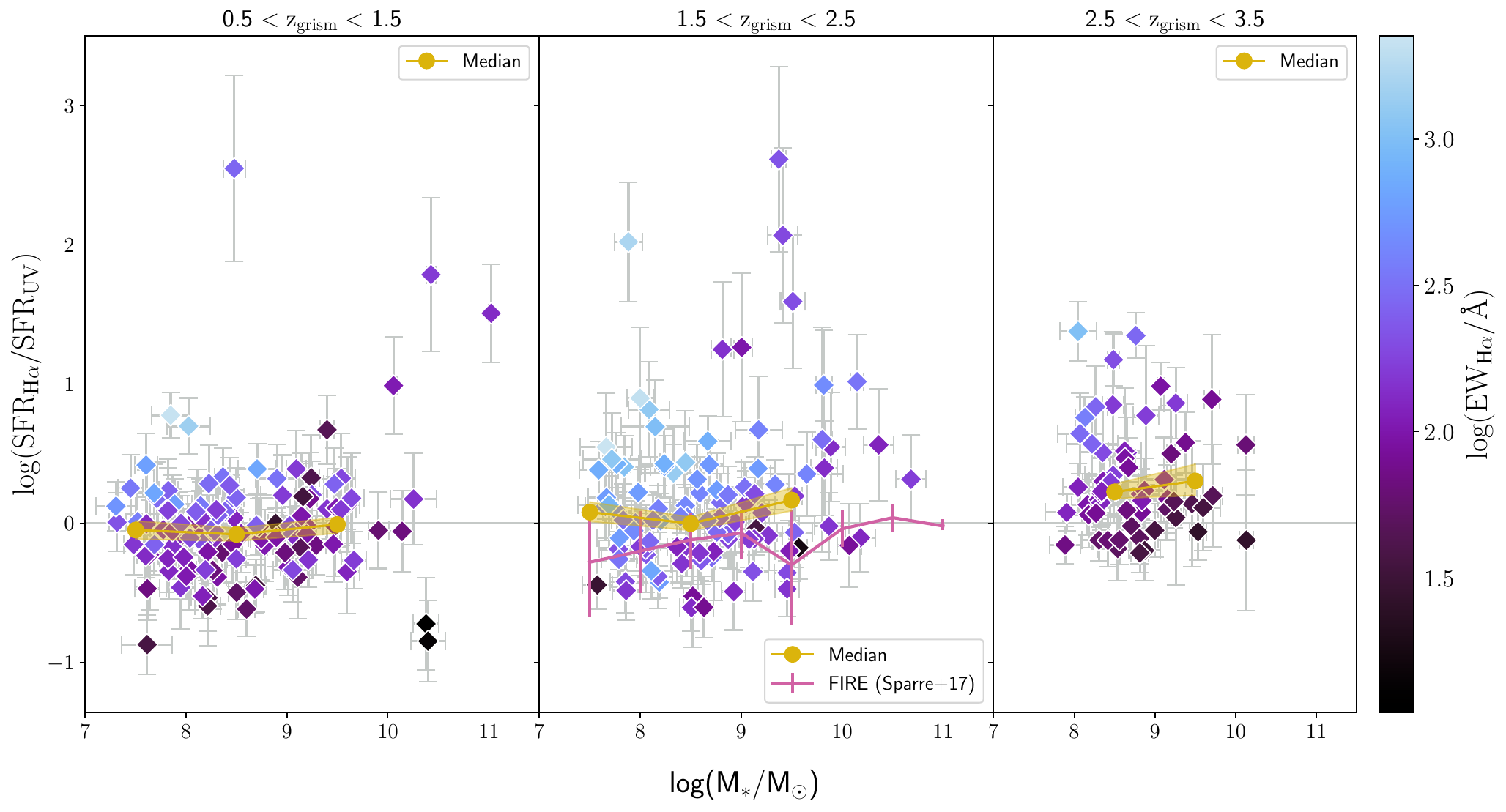}
\caption{The ratio between $\mathrm{SFR_{\Ha}}$ and $\mathrm{SFR_{UV}}$ for our sample as a function of mass in three redshift bins. Galaxies with large \Ha\ EWs are denoted by lighter colors, as indicated by the colorbar, and vice versa. Forecasts from the FIRE simulations \citep{Sparre_2017} regarding the relationship between $\mathrm{\log(SFR_{\Ha}/SFR_{UV})}$ and stellar mass at $z\sim2$ are visible in pink within the middle panel. A grey horizontal line in each panel indicates where $\mathrm{SFR_{\Ha}}$ and $\mathrm{SFR_{UV}}$ are equivalent. Galaxies that have a $\mathrm{\log(SFR_{\Ha}/SFR_{UV})}$ ratio consistent with zero (within a $1\sigma$ uncertainty) have had a constant star-formation rate over the past $\gtrsim$10 Myrs, whereas galaxies that lie above this line have experienced a star-formation ``burst'' in the past $\sim$10~Myr.  Galaxies below the line, on the other hand, have experienced a decline in star formation for $\gtrsim$10~Myr. 
\label{fig:balmer-100_ratio}}
\end{figure*}

Whereas the scatter in the \Ha-- and UV--based SFMSs only informs us about our galaxy sample's relative star formation variations in discrete bins of redshift and stellar mass on 10~Myr versus 100~Myr timescales, we can analyze the relative star-formation intensity on these different timescales for each individual galaxy in our sample by determining the ratio between its \Ha\ SFR and UV SFR, $\mathrm{\log(SFR_{\Ha}/SFR_{UV})}$ \citep{Faisst_2019}. Depicted in Figure \ref{fig:balmer-100_ratio} are the observational $\mathrm{\log(SFR_{\Ha}/SFR_{UV})}$ values we calculate for our sample galaxies, which are represented by diamonds, along with errors propagated from the individual errors on $\mathrm{\log(SFR_{\Ha})}$ and $\mathrm{\log(SFR_{UV})}$ in grey. To estimate the fraction of galaxies with $\mathrm{\log(SFR_{\Ha}/SFR_{UV})} > 0$, which indicates that they are actively forming stars \citep{Cole_2025}, we once again subdivide our sample into nine $\Delta z = 1$, $\Delta \mathrm{\log(M_*/M_{\odot}}) = \mathrm{1 \ dex}$ bins and perform a Monte Carlo simulation with 400 iterations on the $\mathrm{SFR_{\Ha}/SFR_{UV}}$ ratios corresponding to the galaxies in each bin. More specifically, during each iteration, we generate a Gaussian distribution that has a standard deviation equivalent to the uncertainty on $\mathrm{\log(SFR_{\Ha}/SFR_{UV})}$ for each galaxy and draw a random value from that distribution, which we then add to the galaxy's observed $\mathrm{\log(SFR_{\Ha}/SFR_{UV})}$ ratio. Subsequently, we determine the fraction of galaxies that satisfy the condition $\mathrm{\log(SFR_{\Ha}/SFR_{UV})} > 0$ within each bin of redshift and stellar mass containing at least 15 galaxies, and we record the median of the 400 total fractions, as well as uncertainties orginating from the 16th and 84th percentiles of these fractions. At [$0.5 < \mathrm{z_{grism}} < 1.5$, $1.5 < \mathrm{z_{grism}} < 2.5$], the percentage of galaxies with $\mathrm{\log(SFR_{\Ha}/SFR_{UV})} > 0$ ranges from [$45^{+7}_{-7}\%, 59^{+5}_{-9}\%$] in our lowest stellar mass bin, $7 < \mathrm{log(M_*/M_{\odot}) < 8}$, to [$45^{+7}_{-7}\%, 55^{+4}_{-7}\%$] in our highest stellar mass bin, $8 < \mathrm{log(M_*/M_{\odot}) < 9}$. There are only two stellar mass bins at $2.5 < \mathrm{z_{grism}} < 3.5$ with a sufficient number of galaxies available to measure this percentage: in the $8 < \mathrm{log(M_*/M_{\odot}) < 9}$ stellar mass bin, we find that $62^{+3}_{-27}\%$ of galaxies meet the condition $\mathrm{\log(SFR_{\Ha}/SFR_{UV})} > 0$, and likewise $74^{+5}_{-5}\%$ in the $9 < \mathrm{log(M_*/M_{\odot}) < 10}$ stellar mass bin. Simultaneously, we record the median $\mathrm{SFR_{\Ha}/SFR_{UV}}$ value for the binned galaxies in each iteration. The gold diamonds in Figure \ref{fig:balmer-100_ratio} symbolize the median value of the resulting median $\mathrm{SFR_{\Ha}/SFR_{UV}}$ distribution from the Monte Carlo simulation, and the shaded gold regions indicate the 16th and 84th percentiles of this distribution. 

\medskip 

\section{Discussion\label{sec:dis}}

\subsection{Comparisons of Our-Best Fit SFMSs to Previous Surveys\label{subsec:lit}}

In the bottom row of Figure \ref{fig:sfms}, we showcase the best-fit \Ha-- and UV--based SFMSs from \texttt{linmix} for our sample in three redshift bins alongside SFMSs from previous photometric and spectroscopic surveys alike, including the UV-- and IR--based SFMSs reported by \citet{Whitaker_2014}, the \Ha\ SFMSs reported by \citet{Shivaei_2015} and \citet{Clarke_2025}, and the SED--derived, 100 Myr--averaged SFMSs from \citet{Merida_2026}. Additionally, in each redshift bin we overlay SFMSs involving SFRs averaged over 100 Myr from mock galaxies in eight, deep light-cones that cover $132''$ each, and that \citet{Yung_2022} constructed using the Santa-Cruz semi-analytical model (SC SAM). The 50th percentile values of the 100 Myr--averaged SFMS from the SC SAM light-cones are represented by a solid blue line, and light-blue shading surrounding this solid line delineates the 16th and 84th percentiles of the SFMS.

Starting in the bottom-left panel of Figure~\ref{fig:sfms}, we juxtapose our $0.5 < \mathrm{z_{grism}} < 1.5$, \Ha--based SFMS with the $0.5 < z < 1$ and $1 < z < 1.5$ SFMSs that \citet{Whitaker_2014} established using UV and IR photometry of the CANDELS survey area above a stellar mass threshold of $\mathrm{\log(M_*/M_{\odot}) = (8.4, 8.8)}$. Given the $\mathrm{0.48^{+0.04}_{-0.03} \ dex}$ intrinsic scatter of our \Ha--based SFMS (see Table \ref{tab:sfms_scatter}), our results are consistent with those from \citet{Whitaker_2014} above their respective stellar mass limits. Below $\mathrm{\log(M_*/M_{\odot}) \sim (8.4, 8.8)}$, our \Ha-- based SFMS diverges from the extrapolated \citet{Whitaker_2014} relation due to differing assumptions in the shape of the SFMS: \citet{Whitaker_2014} assumes that the SFMS possesses a broken power-law relationship, whereas we opt for a linear model. 

With respect to our $0.5 < \mathrm{z_{grism}} < 1.5$, UV--based SFMS, the slope, intercept, and intrinsic scatter all align closely with those of the $1 < z < 1.5$ SFMS from \citet{Merida_2026}, who derive the 100 Myr--averaged SFRs included in their SFMSs from SED modeling of multiple archival space-based and ground-based photometric surveys. We note that here and throughout the rest of this work, we refer only to the SFMSs which \citet{Merida_2026} determined assuming symmetrical errors on stellar mass and star-formation rates, as we likewise assume in our analysis. \citet{Merida_2026} is able to probe the UV--based SFMS down to remarkably low stellar masses due to the depth of the photometric datasets involved in their analysis, achieving an 80\% stellar mass limit of $\mathrm{log(M_*/M_{\odot})} = 7.6$ at $z=1-1.5$ and $\mathrm{log(M_*/M_{\odot})} = 8.3$ at $z=3-4$. It is noteworthy that our deep spectrosocpic NIRISS observations reach a comparable completeness level, with an 80\% empirical stellar mass limit of $\mathrm{log(M_*/M_{\odot})} = 7.1$ at $z=1-1.25$, and $\mathrm{log(M_*/M_{\odot})} = 7.8$ at $z=3.25-3.5$. 

At $0.5 < \mathrm{z_{grism}} < 1.5$, as well as our two higher redshift bins, we observe that the 100 Myr--averaged SFMSs from the SC SAM exhibit a nonlinear morphology that  renders the comparison between them and our linear, best-fit SFMSs less straightforward than observationally-determined SFMSs from the literature. In general, though, we find that our \Ha-- and UV--based SFMSs are generally consistent with the 50th percentile of the 100 Myr--averaged SFMS from the SC SAM within our lowest redshift bin when the intrinsic scatters of each are taken into account. For the 100 Myr--averaged SFMSs from the SC SAM, we define their intrinsic scatter as half the difference between their 84th and 16th percentiles.

Subsequently, in the bottom-middle panel of Figure~\ref{fig:sfms}, we compare our $1.5 < \mathrm{z_{grism}} < 2.5$, \Ha--based SFMS to the UV-- and IR--based SFMS from \citet{Whitaker_2014} at $1.5 < z < 2.5$, and to the \Ha\ SFMSs from \citet{Shivaei_2015} at $1.37 < z < 2.61$ and \citet{Clarke_2025} at $1.4 < z < 2.7$. Above a limiting stellar mass of $\mathrm{log(M_*/M_{\odot})} = 9.25$, where both our sample and that of \cite{Whitaker_2014} are complete, our \Ha--based SFMS is consistent with the UV-- and IR--based SFMS from \citet{Whitaker_2014} at $1.5 < z < 2.5$ (we note that here, we have averaged the SFMSs at $1.5 < z < 2$ and $2 < z < 2.5$ from \citealt{Whitaker_2014}) within its $\mathrm{0.53^{+0.05}_{-0.05} \ dex}$ intrinsic scatter (see Table \ref{tab:sfms_scatter}). 

Relative to the \Ha\ SFMS from \citet{Shivaei_2015}, our \Ha--based SFMS exhibits a steeper slope $\mathrm{\beta}$ by $\Delta\beta \approx 0.4$, an intercept that is lower by $\approx 0.5 \ \mathrm{dex}$, and an intrinsic scatter that is higher by $\approx 0.2 \ \mathrm{dex}$. The substantially wider stellar mass range covered by our sample could easily account for these offsets. As for the \Ha\ SFMS from the JADES and AURORA surveys \citep{Clarke_2025}, its intercept agrees fairly well with that of our \Ha--based SFMS, with an offset of $\lessapprox 0.25 \ \mathrm{dex}$, but its slope and intrinsic scatter are both lower by $\approx 0.2 \ \mathrm{dex}$.

Conversely, the slope of our UV-based SFMS is $\Delta\beta$ $\approx$ 0.2 steeper than the SED-derived, 100-Myr averaged SFMS reported by \citet{Merida_2026} at $1.5 < z < 2.5$, which we once again have averaged into a single redshift bin from their SFMSs at $1.5 < z < 2$ and $2 < z < 2.5$. In addition, we find a lower intrinsic scatter for our UV--based SFMS by $\approx 0.2 \ \mathrm{dex}$. We suspect that the discrepancy in slope is a result of the far larger sample size of \citet{Merida_2026} at the higher stellar-mass relation of the SFMS. Qualitatively, our UV--based SFMS and the SED--derived, 100--Myr averaged SFMS of \citet{Merida_2026} agree in the stellar mass regime where our SFMS is well sampled.

Lastly, in the bottom-right panel of Figure~\ref{fig:sfms}, we contrast the results of our $2.5 < \mathrm{z_{grism}} < 3.5$, \Ha--based SFMS (which is solely estimated from our \Hbeta--line measurements at these redshifts; see Section~\ref{subsec:sfr} for further details) to the \Ha\ SFMS reported by \citet{Clarke_2025} at $2.7 < z < 4$. As in our intermediate redshift bin, the $2.5 < z_\mathrm{grism} < 3.5$, \Ha--based SFMS lies $< 1 \ \mathrm{dex}$ above the \Ha\ SFMS from \citet{Clarke_2025} at the latter study's stellar mass completeness limit of $\mathrm{log(M_*/M_{\odot})} = 8.5$, although the intrinsic scatter they measure is comparable to ours (see Table \ref{tab:sfms_scatter}). Likewise, the intercept of our UV--based SFMS is higher by $\simeq 0.5$~dex than that of the UV--based SFMS at $2.5 < z < 3$ from \citet{Merida_2026}, but the slope and intrinsic scatter agree. 

One explanation for the higher intercepts of our $2.5 < \mathrm{z_{grism}} < 3.5$ SFMSs compared to the \Ha\ SFMS from \citet{Clarke_2025} and the SED-derived, 100-Myr averaged SFMS from \citet{Merida_2026} is that the SED modeling adopts different assumptions in the construction of star-formation histories.  \citet{Clarke_2025} implemented the SED modeling code \texttt{PROSPECTOR} \citep{Leja_2019, Johnson_2021} in their analysis, which places more prior density on earlier episodes of star-formation.  According to \citet{Leja_2019}, \texttt{PROSPECTOR} yields stellar masses of up to $0.2 \ \mathrm{dex}$ higher than other SED modeling codes, particularly at stellar masses of $\log(M_*/M_{\odot}) < 9$, due to assumptions it makes to account for outshining. \citet{Merida_2026} performed SED modeling using \texttt{Dense Basis} \citep{Iyer_2019}, which likewise tends to overestimate the stellar masses of low-mass galaxies in comparison to various other SED modeling codes like \texttt{cigale}. 

In the case of the $2.5 < \mathrm{z_{grism}} < 3.5$, \Ha--based SFMS specifically, we note that there are numerous galaxies in this redshift bin that simultaneously have \Ha--based SFRs with $S/N < 3$ and lie \textit{above} our formal emission-line detection threshold ($\mathrm{\log(SFR_{\Ha,lim})} \sim 0$), suggesting that this threshold may be underestimated at higher redshifts. We also suspect that this threshold may be lower than that of \citet{Clarke_2025}, who report galaxies with $\mathrm{S/N > 3}$ for $\mathrm{\log(SFR_{\Ha})} < 0$ at $2.7 < z < 4$. If truly present, a difference between our $\mathrm{\log(SFR_{\Ha})}$ detection thresholds would likely contribute significantly to our higher SFMS intercept measurement. Additionally, because we implement $S/N > 3$ cuts for $\mathrm{\log(SFR_{\Ha})}$ in our analysis of the \Ha--based SFMS, there may be an emission-line bias that shifts our \Ha--based SFMS, especially at lower stellar masses. The higher $5\sigma \ \mathrm{\log(SFR_{\Ha})}$ limits displayed in the top and middle rows of Figure \ref{fig:sfms} provide evidence in support of this bias. Previous spectroscopic studies such as \citet{Shivaei_2015}, \citet{Sanders_2018}, and \citet{Papovich_2022} have observed slight $\mathrm{SFR_{\Ha}}$ biases as well; in the case of the CLEAR survey at $1.1 < z < 2.3$ \citep{Papovich_2022}, for instance, emission-line $S/N$ cuts involved in their analysis led to a bias of $\simeq 0.2 \ \mathrm{dex}$ in comparison to the 3D-HST survey.  This accounts for, at least in part, the offset between our results and \citet{Clarke_2025}.  

\subsection{Contrasting the SFMS Scatter to the Literature and Theory}\label{subsec:discussion_bins}

In Figure \ref{fig:scatter}, we present the scatter in the \Ha-- and UV--based SFMSs that we determined using two different methods, the details of which are discussed in Section \ref{sec:results}. Measured as a function of redshift alone, this scatter (i.e., the intrinsic scatter yielded by \texttt{linmix}) is denoted by shaded regions, and as a function of both redshift and stellar mass (which we refer to as $\mathrm{\sigma_{NMAD} \ log(SFR_{corr})}$ in Section \ref{subsec:results_bins}), this scatter is represented as triangles for \Ha\ and squares for UV. We observe an overall decrease in the intrinsic scatter of the \Ha--based SFMS determined using \texttt{linmix} from $0.48^{+0.04}_{-0.03} \ \mathrm{dex}$ at $0.5 < z_\mathrm{grism} < 1.5$ to $\sim 0.35^{+0.04}_{-0.04} \ \mathrm{dex}$ at $2.5 < z_\mathrm{grism} < 3.5$. However, there is no net change in the intrinsic scatter of our UV--based SFMS. \citet{Clarke_2025}, meanwhile, find no appreciable decrease in the intrinsic scatter of both their \Ha-- and UV--based SFMSs until $z \sim 6$. The difference we observe in the evolution of the \Ha--based SFMS's intrinsic scatter is likely driven by our $\approx 0.5 - 1 \ \mathrm{dex}$ lower mass completeness threshold over a shared redshift range of $1 < z \lessapprox 2.5$, which may cause our analysis to be less biased against lower-mass galaxies with a wider range of SFRs. As for our UV--based SFMS, the lack of evolution in its scatter differs slightly from the findings of \citet{Merida_2026}, who report a small, $\approx 0.1 \ \mathrm{dex}$ decline in the scatter from $1 < z < 1.5$ to $3 < z < 4$, but we consider this difference insignificant compared to the uncertainties. With regard to the scatter we measure as a function of both mass and redshift, we find that within each redshift bin, the scatter in the \Ha--based SFMS and UV--based SFMS do not appreciably change as a function of stellar mass, and, in the case of the former, are consistent with their respective intrinsic SFMS scatters. 

Intriguingly, we observe that the intrinsic scatter in our \Ha\ SFMS is consistently higher than that of our UV--based SFMS by $\mathrm{\simeq 0.2-0.4 \ dex}$ except in our highest, $2.5 < \mathrm{z_{grism}} < 3.5$ redshift bin, where the intrinsic scatters of the SFMSs are consistent within their uncertainties.  We note as well that at $z_\mathrm{grism} > 2.5$, we are reliant on \Hbeta--emission and SED-based dust corrections alone to infer the \Ha--based SFMS (see Section~\ref{subsec:sfr} above). The higher scatter we observe in the \Ha--based SFMS also generally persists when we measure the scatter in our \Ha-- and UV--based SFMSs as a function of both stellar mass and redshift. Because of this, at least at $0.5 < z_\mathrm{grism} < 2.5$, where our observations best probe \Ha\ emission, there appears to be greater variability in our sample's star-formation histories on short, 10-Myr timescales.

Quantitatively, our results contradict predictions from the FIRE simulations at $z\sim2$ \citet{Sparre_2017} because the difference between the stellar mass-- and redshift--dependent scatter in our $\mathrm{1.5 < z_{grism} < 2.5}$, \Ha-- and UV--based SFMSs ranges from $\mathrm{0.3-0.55}$~dex, whereas for the FIRE simulations at $z\sim2$, this disparity is merely $\simeq 0.05 - 0.1$ dex. Furthermore, the FIRE simulations only agree with the scatter in our $1.5 < \mathrm{z_{grism}} < 2.5$, \Ha--based SFMS within a $1\mathrm{\sigma}$ uncertainty at the highest stellar masses ($\mathrm{\log(M_*/M_{\odot})} > 9$); at lower stellar masses, the scatter in the FIRE simulations is up to $\approx0.25~$dex lower than our observed scatter. 

On the contrary, the observational scatter we measure as a function of stellar mass in our UV--based SFMS at $\mathrm{1.5 < z_{grism} < 2.5}$, which ranges from $\approx 0.25-0.3$~dex over $7 < \mathrm{\log(M_*/M_{\odot})} < 10$, is lower than predicted by either the FIRE simulations or the previously discussed lightcones from the SC SAM \citep{Yung_2022}. With regard to the mock galaxies' SFMS in the SC SAM lightcones, we quantify its scatter by dividing the galaxies into four equal stellar mass bins from $8 < \mathrm{\log(M_*/M_{\odot})} < 10$ and then determining half of the difference between the 16th and 84th percentiles of their 100-Myr averaged SFRs. As shown in Figure~\ref{fig:scatter}, the FIRE simulations forecast that, at a fixed redshift of $z\sim2$, the scatter in the UV--based SFMS should be $\approx 0.15$~dex higher than our observed scatter is at $8 < \mathrm{\log(M_*/M_{\odot})} < 10$, and the SC SAM lightcones similarly anticipate a higher scatter in the 100-Myr averaged SFMS than we measure by $\approx 0.15-0.2$~dex over the same stellar mass range. Two explanations exist for these disparities: the first involves a bias in the observed SFRs of our sample, and the second an issue with the feedback recipes adopted by theoretical models. Expanding upon the latter scenario, increasing the feedback strength in theoretical models would decrease the amount of time for which star formation bursts can endure before their gas supply is disrupted or removed by galactic-scale outflows, thereby evoking intense bursts of star formation on 10 Myr timescales that interrupt long, $\gtrsim$100~Myr periods of otherwise constant (or declining) star formation. In effect, this scenario may reproduce for our sample galaxies at $z\sim2$ both the highly stochastic star formation histories we observe on 10 Myr timescales and the smoother star formation histories on 100 Myr timescales that we observe in comparison to theoretical predictions.

\subsection{Burstiness and Duty Cycles}

The deviation of a SFMS galaxy's $\mathrm{SFR_{\Ha}/SFR_{UV}}$ ratio from zero chronicles its star-formation history during the past $\sim10-100$~Myr by illuminating periods of uneventful, steady star-formation ($\mathrm{\log(SFR_{\Ha}/SFR_{UV}}) \simeq 0$), momentous bursts of star formation ($\mathrm{\log(SFR_{\Ha}/SFR_{UV}}) > 0$), and dormant, ``napping'' states of temporarily declining star-formation ($\mathrm{\log(SFR_{\Ha}/SFR_{UV}}) < 0$) \citep{Faisst_2019}. This quantity provides constraints on the duration of star formation bursts that a galaxy experiences, and an estimate of the ``duty cycle'' as well, which is the fraction of time during which a galaxy actively forms stars.  We note that ideally, the ratio between these SFRs should be measured using SFRs that are directly inferred from rest-UV observations rather than our ``UV--based SFRs,'' which are in actuality derived from SED modeling, and we would have done so if such observations were available. However, we do notice a reassuring correlation between $\mathrm{\log(SFR_{\Ha}/SFR_{UV}})$ and \Ha\ equivalent width (EW), an unequivocal tracer of active star formation, such that galaxies with higher $\mathrm{\log(SFR_{\Ha}/SFR_{UV}})$ ratios exhibit higher \Ha\ EWs as well, particularly at low stellar masses. Still, we dedicate particular care to the assessment and implications of these results, and advise the reader to do the same. 

As is evident in Figure \ref{fig:balmer-100_ratio}, the galaxies in our sample have $\mathrm{log(SFR_{\Ha}/SFR_{UV})}$ ratios that are generally distributed around zero at $0.5 < \mathrm{z_{grism}} < 2.5$ regardless of stellar mass, which provides further support for largely constant star formation histories within our sample in the past $\sim 100$~Myr. While we do find that the median $\mathrm{log(SFR_{\Ha}/SFR_{UV})}$ ratio is marginally elevated above zero at $\mathrm{8 < \log(M_*/M_{\odot}) < 10}$ in our highest redshift bin, $2.5 < \mathrm{z_{grism}} < 3.5$, we caution against interpreting this observation as an increase in the frequency of bursts at higher redshifts due to our substantially higher SFR detection thresholds at these redshifts, which we expand upon in Section \ref{subsec:lit}. At $z\sim2$, we consider our median $\mathrm{log(SFR_{\Ha}/SFR_{UV})}$ ratios at varying stellar masses in the context of predictions from the FIRE simulations \citep{Sparre_2017} for the relationship between $\mathrm{log(SFR_{\Ha}/SFR_{UV})}$ and stellar mass. Within 1$\sigma$ uncertainties, our median $\mathrm{log(SFR_{\Ha}/SFR_{UV})}$ ratios and the FIRE projections are concordant at all stellar masses. 

With respect to the fraction of galaxies in our sample with $\mathrm{log(SFR_{\Ha}/SFR_{UV})} > 0$, we recognize that there is no substantial evolution at fixed stellar mass over $0.5 < \mathrm{z_{grism} < 2.5}$ considering the uncertainties on these fractions. We do not consider our highest redshift bin in the assessment of this fraction's evolution due to the reasons discussed above. Overall, we find that roughly half the galaxies in our sample are actively forming stars ($\mathrm{\log(SFR_{\Ha}/SFR_{UV}}) > 0$) at all stellar masses over $0.5 < \mathrm{z_{grism} < 2.5}$, which suggests that the duty cycle of star formation in each redshift bin is equivalent to $\approx 50\%$ and that galaxies undergo active star formation for around half of the time elapsed during each bin according to Equation 9 from \citet{Cole_2025}. These findings are a natural consequence of our median $\mathrm{log(SFR_{\Ha}/SFR_{UV})}$ ratios' general consistency with zero over all stellar masses within this redshift range. 

At higher redshifts ($\mathrm{z = 4.5 - 9}$), \cite{Cole_2025} constructed a similar metric, $\mathrm{\log(SFR_{10}/SFR_{100}})$, using SFRs that were averaged over 10~Myr and 100~Myr, respectively. Using the fraction of galaxies with $\mathrm{\log(SFR_{10}/SFR_{100}}) > 0$, \cite{Cole_2025} concluded that the star formation histories of low-mass CEERS galaxies ($\mathrm{\log(M_*/M_{\odot}) \lesssim 9}$) are likewise characterized by duty cycles of $\approx 50\% \pm 10\%$, which is qualitatively consistent with the duty cycles of our NGDEEP galaxies. However, \cite{Perry_2025}'s spectroscopic analysis of the ratio between \Ha\ and UV luminosities for low-mass galaxies ($\mathrm{\log(M_*/M_{\odot}) < 8.5}$) in the CEERS and RUBIES surveys at $\mathrm{z = 1-2.5}$ with JWST/NIRSpec revealed duty cycles that are up to $\approx20-25\%$ higher than for NGDEEP galaxies with stellar masses of $\mathrm{7 < \log(M_*/M_{\odot}) < 9}$ in our $0.5 < \mathrm{z_{grism}} < 1.5$ redshift bin, yet are consistent with our NGDEEP galaxies at $1.5 < \mathrm{z_{grism}} < 2.5$.  Collectively, the results of our work and of \citep{Perry_2025} tentatively suggest that the duty cycles of low-mass galaxies may begin to decline at $z \lesssim 1$, implying that galaxies at lower redshift spend less time in bursty stages, although there is not a sufficient number of NGDEEP galaxies available at $0.5 < \mathrm{z_{grism}} < 1$ for us to explore this eventuality with any certainty.
%
%This suggests that for these galaxies have star-formation histories with bursts with amplitudes and durations similar to those predicted by the models. However, our results show high scatter, $\simeq$ 0.5--0.7~dex in the \Ha-based SFMS at lower masses, $\log M/M_\odot = 8-9$ compared to FIRE, $\simeq$ 0.4~dex.  This implies bursts may be stronger on short timescales in the data compared to these predictions.  Our observed scatter continues to be high, 0.4--0.7~dex at lower stellar masses, $log M/M_\odot$=7--8, implying there is no evidence for a decline in ``burstiness''. 

\section{Summary}

In this paper, we probe the star-formation activity of NGDEEP galaxies with stellar masses of $\mathrm{7 \lesssim \log(M_*/M_{\odot}) \lesssim 11}$ at $0.5 < \mathrm{z_{grism}} < 3.5$ through \Ha--based SFRs measured from JWST/NIRISS observations of the HUDF and UV--based SFRs derived via SED-fitting of ancillary legacy imaging from HST/ACS and JWST/NIRCam. Leveraging these deep observations, we place a special emphasis on low-mass ($\mathrm{\log(M_*/M_{\odot})} \lesssim 9$) galaxies in our analysis of the NGDEEP sample's star-forming main sequence and star formation variability.  

Bayesian linear regression modeling of our sample's $\mathrm{SFR-M_*}$ distributions with \texttt{linmix}, which we perform for each SFR tracer, reveals the intrinsic scatter within the \Ha-- and UV--based SFMS. Over $0.5 < \mathrm{z_{grism}} < 3.5$, the intrinsic scatter in the \Ha--based (UV--based) SFMS ranges from $\approx 0.35\ \mathrm{to} \ 0.5 \ \mathrm{dex}$ ($\approx 0.15\ \mathrm{to} \ 0.3 \ \mathrm{dex}$). Although the intrinsic scatters in each SFMS agree at $\mathrm{z_{grism}} > 2.5$, a $\approx 0.2-0.4 \ \mathrm{dex}$ difference exists between them at $\mathrm{z_{grism}} < 2.5$, signifying that our NGDEEP sample experiences more star formation variability on $\lesssim 10$~Myr timescales than $\sim100~$Myr timescales. 

By measuring the scatter within each SFMS as a function of stellar mass using a Monte Carlo simulation, we find that the scatter in our \Ha--based SFMS is consistently high ($\approx 0.5 - 0.7 \ \mathrm{dex}$) across our sample's stellar mass range ($7 \lesssim \mathrm{\log(M_*/M_{\odot})} \lesssim 10$) at $1.5 < \mathrm{z_{grism}} < 2.5$, even in comparison to the $\approx 0.4 - 0.45 \ \mathrm{dex}$ \Ha\ scatter predictions from the FIRE simulations at $z\sim2$. Due to the low scatter ($\approx 0.2-0.25$~dex) we observe in our UV--based SFMS at the same redshifts and stellar masses, however, the intense star bursts of star formation suggested by our high \Ha\ scatter measurements appear to be brief interruptions in the past $\sim 100$~Myr of our sample galaxies' otherwise smooth star-formation histories. In the future, theoretical predictions may need to adjust current prescriptions for feedback to reproduce the intense bursts of star formation that are indicated by our high observed scatter measurements. 

Additionally, we assess the fraction of galaxies in our sample that are actively forming stars through the ratio between their \Ha-- and UV--based SFRs. At $\mathrm{z_{grism}} < 2.5$, we find that the median $\mathrm{log(SFR_{\Ha}/SFR_{UV})}$ ratio is overall consistent with zero across our sample's stellar mass range, providing further evidence of smooth star formation histories in our sample. In effect, this observation suggests as well that about half of our sample is actively forming stars ($\mathrm{log(SFR_{\Ha}/SFR_{UV})} > 0$), which yields duty cycle estimates for our $\mathrm{z_{grism}} < 2.5$ NGDEEP sample galaxies of approximately $50\%$.

For our final remarks, we emphasize to the reader that our NGDEEP survey consists of a single NIRISS pointing. Although the depth of our dataset supplies us with exquisite spectra and emission-line fluxes for exceptionally low-mass galaxies (considering their redshifts), far larger samples with comparable depths are necessary to reduce the fairly substantial uncertainties in our SFMS scatter measurements and to pinpoint the magnitude of the differences that exist between theoretical predictions and observations. With the help of new and future WFSS surveys from space-based telescopes that encompass much vaster swaths of sky, these challenges will soon undoubtedly be overcome. 

\begin{acknowledgments}
We acknowledge the invaluable contributions of our colleagues in the NGDEEP collaboration and  those involved in the JWST mission to this work, for which we are very grateful. This work benefited from support from the George P. and Cynthia Woods Mitchell Institute for Fundamental Physics and Astronomy at Texas A\&M University. 
CP thanks Marsha and Ralph Schilling for generous support of this research. 
Additionally, this work is based on observations made with the NASA/ESA/CSA James Webb Space Telescope through the Space Telescope Science Institute, which is operated by the Association of Universities for Research in Astronomy, Inc., under NASA contract NAS 5-03127 for JWST. Specifically, the NGDEEP Survey observations are associated with JWST program \#2079. CB thanks Justin W. Cole for insightful discussions about star-formation histories, which have without a doubt improved the quality of this work. LYAY acknowledges support from the Taiwanese Ministry of Education (MoE) Yushan Fellowship (MOE-115-YSFMS-0002-005-P1). 
\end{acknowledgments}

\bibliography{sfms}{}
\bibliographystyle{aasjournal}

\end{document}